\documentclass[%
reprint,
amsmath,amssymb,
aps,
]{revtex4-2}

\usepackage{graphicx}
\usepackage{dcolumn}
\usepackage{bm}
\usepackage{hyperref}
\hypersetup{
	colorlinks=true,
	linkcolor={black},
	citecolor={black},
	urlcolor={black},
}

\usepackage{enumitem}
\usepackage{notes2bib}
\bibnotesetup{
	note-name = ,
	use-sort-key = false
}

\graphicspath{{figures/}}

\begin{document}

	\title{Nonequilibrium model of short-range repression in gene transcription regulation}
	
	\author{F.E. Garbuzov}
	\author{V.V. Gursky}%
	\email{gursky@math.ioffe.ru}
	\affiliation{%
		Ioffe Institute, 26 Polytekhnicheskaya, St. Petersburg 194021, Russia
	}%
	
	\date{November 10, 2021}
	
	\begin{abstract}
		Transcription factors are proteins that regulate gene activity by activating or repressing gene transcription. A special class of transcriptional repressors operates via a short-range mechanism, making local DNA regions inaccessible to binding by activators and, thus, providing an indirect repressive action on the target gene. This mechanism is commonly modeled assuming that repressors interact with DNA under thermodynamic equilibrium and neglecting some configurations of the gene regulatory region.
		We elaborate a more general nonequilibrium model of short-range repression using the graph formalism for transitions between gene states, and we apply analytical calculations to compare it with the equilibrium model in terms of the repression strength and expression noise. In contrast to the equilibrium approach, the new model allows us to separate two basic mechanisms of short-range repression. The first mechanism is associated with the recruiting of factors that mediate chromatin condensation, and the second one concerns the blocking of factors that mediate chromatin loosening. The nonequilibrium model demonstrates better performance on previously published gene expression data obtained for transcription factors controlling {\em Drosophila} development, and furthermore, it predicts that the first repression mechanism is the most favorable in this system. The presented approach can be scaled to larger gene networks and can be used to infer specific modes and parameters of transcriptional regulation from gene expression data.
	\end{abstract}
	
	\maketitle
	
	
	\section{Introduction}
	
	Gene regulation is crucial for understanding many biological processes~\cite{Struhl1999}. There is a growing need for quantitative models of gene regulation to describe the increasing amount of experimental data obtained with various data acquisition techniques~\cite{Li:2011kza,Coulon:2013bz,Ferraro:2016im}.
	A variety of data-driven modeling approaches have been developed to describe the spatio-temporal dynamics of the expression products (mRNA and proteins) in gene networks operating in various model organisms, including Boolean models, continuous models based on reaction-diffusion equations, models based on statistical thermodynamics, and stochastic models~\cite{deJong:2002ft, Paulsson:2005da, Bintu2005, Bintu2005a, Jaeger2009, Ay2011, Munsky:2012ie, Samee2014, Samee2015, Martinez:2013dl}. 
	
	Thermodynamics-based models are the next step after phenomenological gene expression models, providing enough details for calculating gene expression levels from a DNA sequence while reducing the potentially huge number of free parameters by imposing equilibrium constraints on gene regulation. Transcriptional regulation is performed by transcription factors (TFs), which bind multiple energetically preferable binding sites within the DNA regulatory regions and form specific molecular configurations of the regulatory DNA-protein complexes. The thermodynamics-based models calculate the probabilities of these configurations under thermodynamic equilibrium conditions~\cite{Buchler:2003ed}. 
	Relating the regulatory DNA sequence to the target gene activity, some models assume that certain DNA-bound TFs (activators) facilitate the association of co-factors that help to surpass an energy barrier to initiate transcription~\cite{Reinitz2003, Janssens2006}. Other models consider the basal promoter of the target gene as a binding site for the basal transcriptional machinery, considering this binding site in addition to the TF binding sites from the regulatory sequence, and they calculate the expression rate as proportional to the fractional occupancy of the promoter~\cite{He2010}.
	
	Thermodynamic models based on the assumption of thermodynamic equilibrium, which requires that there is no net macroscopic flow of energy, are widely used to describe gene regulation~\cite{He2010, Fakhouri2010, Kozlov2014, Jaeger2011}. However, gene expression and its regulation implicate several energy dissipating mechanisms, such as reorganization of chromatin, the assembly and movement of nucleosomes, the post-translational modifications of histones, RNA synthesis, etc.~\cite{Gunawardena2020}. These mechanisms consume energy which drives the regulation of gene expression away from equilibrium. For example, single-molecule data show that nonequilibrium mechanisms driven by transcription initiation rule out simple operator occupancy models of gene regulation in living \textit{E. coli} cells~\cite{Hammar-2014}. Several attempts were made to build models that take into account nonequilibrium mechanisms~\cite{Ghosh2015, Horowitz2017}, and a unifying graph formalism for the nonequilibrium Markovian chain based modeling was developed~\cite{Ahsendorf2014, Schnakenberg1976}, which we actively use in our work.
	
	
	TFs that regulate gene activity split into activators and repressors of target genes. Although the molecular mechanisms by which activators and repressors regulate gene activity in metazoa, and in particular in such model organisms as \textit{Drosophila}, are not fully understood, evidence suggests that DNA-bound activators recruit co-factors, or ``adaptor factors''~\cite{Reinitz2003}, that facilitate the binding of the basal transcriptional machinery to the promoter via an enhancer--promotor loop and, thus, they initiate transcription~\cite{Frietze-2011, Small-2020}. DNA-bound repressors recruit co-factors that induce histone deacetylation, thereby facilitating chromatin compaction~\cite{Courey2001, Li2011, Chambers2017}. Repressive TFs are classified into long- and short-range repressors, depending on the range of chromatin compaction provided by their associated co-factors~\cite{Gray1996, Courey2001}. Short-range co-factors lead to the local deacetylation of nucleosomes and chromatin condensation in the vicinity of a bound repressor. Thereby, short-range repressors displace neighboring activators and indirectly inhibit gene activation. \textit{Drosophila} CtBP is an example of such a co-factor, associated with the short-range repressors Giant, Kr\"{u}ppel, Knirps, and Snail~\cite{Mannervik1999}. It is assumed that long-range co-repressors are triggered locally, but polymerize and spread along chromatin, leading to repression of a large chromosomal locus that may comprise distant enhancers and gene promoters~\cite{Courey2001}. Groucho is an example of a long-range co-repressor associated with the long-range repressor Hairy in early \textit{Drosophila} segmentation~\cite{Li2011, Chambers2017}. It was suggested that short-range repression is important for maintaining the autonomous functioning of multiple enhancers involved in the regulation of \textit{Drosophila} development~\cite{Gray1996}. Short-range repression was implemented in the equilibrium thermodynamic models describing the regulation of developmental genes~\cite{Janssens2006, Fakhouri2010, He2010, Kozlov2014, Kozlov2015, Hoermann2016}.
	
	Current data allow a relationship to be established between the presence of repressors and increased histone density of targeted enhancer regions, but there are no specific details of the putative interaction between the repressor and the co-factor responsible for chromatin compactification~\cite{Courey2001, Li2011, Chambers2017, Small-2020}. We consider two possible scenarios for this interaction, and call we them mechanisms of short-range repression. In the first mechanism, the repressor recruits co-factors that promote local chromatin condensation but does not control the stability of the condensed state. In the second one, the repressor does not mediate the initiation of chromatin condensation but stabilizes this state when it is formed. In this study, we examine these two short-range repression mechanisms, formalizing them in terms of rates of specific processes in a simple model of gene regulation. We show that the equilibrium formalism does not allow distinguishing between the mechanisms at the level of mean mRNA copy number transcribed from the target gene, and the nonequilibrium model must be used for that. 
	
	We formulate a minimal model that can incorporate short-range repression in order to get feasible analytical results. We compare the equilibrium and nonequilibrium versions of the model in terms of the repression strength and gene expression noise, and we demonstrate the advantages of the nonequilibrium approach. Using previously published gene expression data~\cite{Fakhouri2010}, we show that the nonequilibrium model outperforms the equilibrium one and predicts the first mechanism of short-repression as preferable. {
		The main goal of our work is to show that the nonequilibrium model provides biophysical insights into the mechanisms of short-range repression which cannot be obtained from an equilibrium model.}

	\section{Models of transcriptional regulation with short-range repression}
	
	\subsection{Regulation of gene transcription}
\begin{figure}[ht]
		\centering
		\includegraphics[width=.8\linewidth]{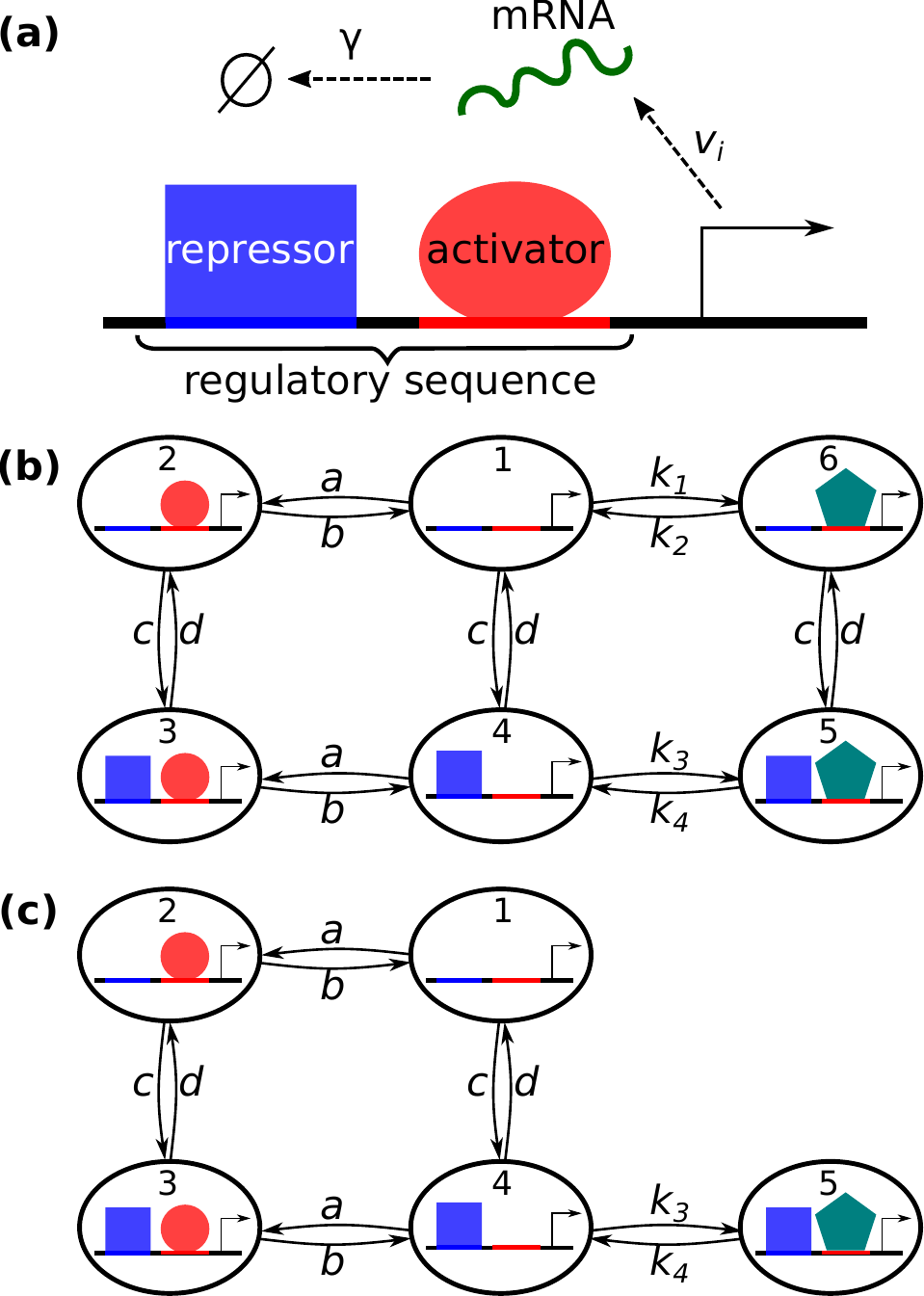}
		\caption{Model of gene expression under the influence of a regulatory region consisting of two binding sites. (a) Schematic of transcription. (b) Full graph of transitions between the regulatory region states (called `gene states' in what follows). (c) Reduced transition graph considered in the equilibrium approximation. Red circles denote bound activators, blue squares denote bound repressors, and teal pentagons denote nucleosomes formed at the activator site, making this site inaccessible for the activator. Edge labels denote rates of transitions.}
		\label{fig:graphs}
	\end{figure}

	We consider a simple regulatory region in the DNA that controls transcription of one gene (Fig.~\ref{fig:graphs}~(a)). The region consists of two transcription factor binding sites, one for an activator and one for a short-range repressor. The regulatory region can be in six different states (Fig.~\ref{fig:graphs}~(b)):
	\begin{enumerate}[label=\arabic*), topsep=0pt, itemsep=-1ex, partopsep=1ex, parsep=1ex]
		\item activator and repressor sites are free,
		\item activator is bound, repressor site is free,
		\item both activator and repressor are bound,
		\item activator site is free, repressor is bound,
		\item activator site is inaccessible (covered by nucleosome), repressor is bound,
		\item activator site is inaccessible, repressor site is free.
	\end{enumerate}
	At each state $i$ ($1\le i\le 6$), messenger RNA (mRNA) is produced on the gene at a rate $v_i$ and degrades at a constant rate $\gamma$ (Fig.~\ref{fig:graphs}~(a)). 
	We assume that the binding and unbinding rates of the activator ($a$, $b$) are independent of whether the repressor is bound or not, and the same rates for the repressor ($c$, $d$) do not depend on the presence of the activator or nucleosome on their site. The repression mechanism in the model is expressed by the fact that the bound repressor increases the occupancy of the activator site by the nucleosome, so the chromatin remodeling rates in the absence ($k_1$, $k_2$) and presence ($k_3$, $k_4$) of repressor are different. The association rates $a$ and $c$ absorb concentrations of the activator ([A]) and repressor ([R]) in the cell, so that 
	\begin{equation} \label{a_c}
		a = a_0 \text{[A]} \text{ and } c = c_0 \text{[R]},
	\end{equation}
	with some proportionality constants $a_0$ and $c_0$ defined by the specificity of the binding sites and the site search kinetics of TFs.
	
	The described model (model based on Figures~\ref{fig:graphs}~(a) and~(b)) is inherently nonequilibrium even in a steady state unless a detailed balance holds. The detailed balance is a fundamental constraint on equilibrium systems which requires the forward probability flow from one state to another to be the same as the backward one. Applying the detailed balance to this model results in the absence of repression. This is because, under this assumption, the probability flow to the state with the closed activator site does not depend on the presence of a repressor on its site, as we will discuss in more detail later. To include the short-range repression mechanism into the equilibrium thermodynamic framework, models were formulated based on a reduced state transition graph, in which the sixth state was omitted (Fig.~\ref{fig:graphs}~(c))~\cite{He2010,Kozlov2014,Kozlov2015}. These models always satisfy the detailed balance in a steady state. In the reduced graph, the bound repressor is a prerequisite for the probability flow to the state with the closed activator site, thus making repression possible under the detailed balance.
	
	The shift from the full graph from Fig.~\ref{fig:graphs}~(b) to the reduced graph from Fig.~\ref{fig:graphs}~(c) is biologically justified only if a local chromatin condensation in the vicinity of the activator site is negligible in the absence of a bound repressor. This assumption should be considered too restrictive, especially in early \textit{Drosophila} segmentation, since the repressor considered in the model is most likely not the exclusive DNA-binding partner of the chromatin remodeling co-factors responsible for the local chromatin compaction. As a developmental process, \textit{Drosophila} segmentation is controlled by multiple enhancers containing multiple binding sites for several short- and long-range repressors~\cite{Small-2020}. These binding sites are tightly packed and show essential overlap~\cite{Cheng2013}, so it is reasonable to regard any given repressive TF as operating on the background of other local DNA-bound repressors that may recruit histone-modifying enzymes in the same local vicinity. Moreover, as we pointed out above, this reduction of the state transition graph historically appeared as a byproduct of the modeling methodology and was not based on biological data.

	In what follows, we investigate a more general model based on the full transition graph (Figures~\ref{fig:graphs}~(a) and~(b)) in the nonequilibrium context, referring to it as the `nonequilibrium model,' and we compare it to the model based on the reduced graph (Figures~\ref{fig:graphs}~(a) and (c)) under the detailed balance, referring to the latter as the `equilibrium model.'
	
	It should be mentioned that the detailed balance is a necessary, but not sufficient condition for the system to be at equilibrium. In a genuine equilibrium system, the ratio of the forward and backward rates, e.g. the rates $a$ and $b$ of transitions between states 1 and 2 in Fig.~\ref{fig:graphs}, comes from the Boltzmann distribution:
	$$ \frac{a}{b} \sim e^{-\Delta W_{12}},$$
	where $\Delta W_{12}$ is the difference in the Gibbs free energy between the two states. If mechanisms associated with energy dissipation participate in transitions between some pairs of connected states, the system is away from equilibrium. We leave the deviation of rate constants outside the scope of our study and consider the equilibrium only in terms of the detailed balance.
	
	\subsection{Gene state probabilities}
	The change in gene state probabilities over time is defined by the Laplacian matrix $L$ of the transition graph as follows:
	\begin{equation}\label{probabilities_eqn}
		\dot{\bm p} = L  \bm p,
	\end{equation}
	where $\bm p(t) = \{p_i(t)\}_{i=1}^m$ is the vector of gene state probabilities, $m$ is the total number of the states ($m=6$ and $m=5$ for the full and reduced graphs, respectively), $L_{ii}$ is the negative sum of all outgoing edge labels from vertex $i$, and $L_{ij}$ is equal to the edge label from vertex $j$ to $i$ if this edge exists and 0 otherwise.
	The steady-state solution $\bm{p}^* = \bm p(t\to\infty)$ of the Eq.~\eqref{probabilities_eqn} belongs to the Laplacian matrix kernel, which is one-dimensional in the case of a strongly connected graph:
	\begin{equation} \label{ker_span}
		\bm{p}^* \in \ker L, \quad \dim\ker L = 1, \quad \ker L = \text{span}\{\bm\rho^*\}.
	\end{equation}
	The probability vector $\bm{p}^*$ is obtained from $\bm{\rho}^*$ by normalization, ensuring that the sum of all vector components is equal to 1.
	
	In the general case, component $\rho^*_i$ of the Laplacian matrix kernel element $\bm\rho^*$ is equal to the sum of products of the edge labels of all spanning trees ($\Theta_i$) rooted at vertex $i$~\cite{Mirzaev2013}:
	\begin{equation}\label{laplace_ker}
		\rho^*_i = \sum_{\theta\in\Theta_i} \bigg( \prod_{j \overset{r}{\to} l \in \theta} r \bigg).
	\end{equation}
	If the graph represents a system that reaches thermodynamic equilibrium, the detailed balance must be satisfied. This means that, for each forward transition from state $i$ to state $j$ with the rate $r_{i\to j}$, there is a backward one with the rate $r_{j\to i}$, and the following equality holds:
	\begin{equation}\label{detailed_balance}
		\rho_j^* = \frac{r_{i\to j}}{r_{j\to i}} \rho_i^*,
	\end{equation}
	which represents the equivalence of the forward and backward probability flows. Using this equation, the stationary probabilities of all states in the equilibrium system from Fig.~\ref{fig:graphs}~(c) can be easily found, and these probabilities depend only on the ratios of forward and backward transition rates (see Eq.~\eqref{stat_weights_eq} in the Appendix). This is not true for the nonequilibrium system from Fig.~\ref{fig:graphs}~(b), where all probabilities depend on each transition rate separately (see Eq.~\eqref{stat_weights_neq} in the Appendix). Application of the detailed balance \eqref{detailed_balance} to the nonequilibrium system yields the equality 
	$$\frac{k_1}{k_2} = \frac{k_3}{k_4},$$
	which means that the bound repressor does not make the activator site less accessible than the unbound repressor state and, therefore, there is no repression in the full graph under the detailed balance assumption.

	\subsection{Master equation and distribution moments for the mRNA copy number}
	A stochastic model of gene transcription can be formulated as the master equation for the whole system, which includes the regulatory region together with the mRNA copy number transcribed from the gene, taking the following form:
	\begin{subequations}\label{master_eqn}
		\begin{eqnarray}
			\dot{\bm{p}}^{(0)} &=& (L-T)\bm{p}^{(0)} + D\bm{p}^{(1)},\\
			\dot{\bm{p}}^{(n)} &=& T\bm{p}^{(n-1)} + (L-T-nD)\bm{p}^{(n)} \nonumber \\
			&&+ (n + 1)D \bm{p}^{(n+1)}, \quad n \geqslant 1,
		\end{eqnarray}
	\end{subequations}
	where $\bm{p}^{(n)}(t) = \{p_i^{(n)}(t)\}_{i=1}^m$ is a vector of gene state probabilities when the system contains $n$ mRNA molecules, $L$ is the Laplacian matrix introduced before, $T = \text{diag}\{v_1, v_2, \dots, v_m\}$ denotes mRNA production, $D = \text{diag}\{\gamma, \gamma, \dots, \gamma\}$ denotes mRNA degradation, and $m$ is the total number of gene states. The model~(\ref{master_eqn}) represents the standard chemical master equations derived from the basic principles~\cite{vanKampen2007}; similar models can be found in many applications~\cite{Brown2013, Munsky2015}.
	
	The total gene state probabilities are equal to the sum of probabilities for all numbers of mRNA:
	$$ \bm{p}(t) = \sum_{n=0}^{\infty} \bm{p}^{(n)}(t), \quad \bm{p}^* = \sum_{n=0}^{\infty} \bm{p}^{(n)}(t\to\infty) $$
	Using the generating function method~\cite{Peccoud1995}, all moments of the mRNA copy number distribution can be found, and we have the following solutions for the steady-state mean and variance:
	\begin{eqnarray} 
		\label{mrna_mean_steady}
		\mu &=&  \frac{\bm{p}^* \cdot \bm{v}}{\gamma}\\
		\label{mrna_variance_steady}
		\sigma^2 &=& \left( \sum_{i=1}^{m} g_i \right) + \mu - \mu^2,
	\end{eqnarray}
	where $\bm v = \{v_i\}$, and $\bm{g} = 2 [(L - 2D)^{-1} T] [(L - D)^{-1} T] \bm{p}^*$.
	It follows from~\eqref{mrna_mean_steady} that the mean mRNA copy number in the equilibrium system, similarly to the state probabilities, depends only on the ratios of transition rates. The mRNA variance in both the equilibrium and nonequilibrium systems depends on each transition rate separately.

	\subsection{Analysis and comparison of the models}
	
	Previous experimental studies showed that the short-range repression by the TF Knirps can be associated with the increased histone density of targeted regulatory regions in the \textit{Drosophila} genome, and two mechanisms were proposed to explain this association~\cite{Li2011}. One mechanism suggests that Knirps recruits factors mediating chromatin condensation, while the second mechanism invokes blocking of proteins responsible for the loosening of chromatin.
	In terms of our model, the contribution of the first mechanism to the total repression is defined by the value of rate $k_3$, while the contribution of the second mechanism is defined by the value of rate $k_4$. Therefore, in this section we are focused on studying the dependence of the model predictions for mean expression levels and expression noise on $k_3$ and $k_4$ and whether this dependence is symmetrical. 
	
	For simplicity, in what follows we assume that mRNA is produced at a constant rate $v_A$ when the activator is bound, i.e. the gene is either in state 2 or 3, and in all other states it is produced at a basal rate $v_0<v_A$:
	\begin{equation}
		v_i = \begin{cases}
			v_A & \text{if } i = 2 \text{ or } 3,\\
			v_0 & \text{otherwise.}
		\end{cases}
	\end{equation}
	
	We compare the equilibrium and nonequilibrium models of short-range repression by analyzing a repression factor $C_R$, which represents repression strength via the relative decrease in mean mRNA copy number due to the presence of repressor, as follows:
	\begin{equation} \label{repr_factor}
		C_R = \frac{\mu_0-\mu}{\mu_0 - v_0/\gamma} = 1 - \frac{p_2^* + p_3^*}{p_{0,2}^*},
	\end{equation}
	where $\mu_0$ and $p_{0,2}^*$ denote the stationary mean mRNA copy number and stationary probability of the activated state, respectively, in a system without a repressor ($c = 0$ in Fig.~\ref{fig:graphs}), while $\mu$ and $p_2^* + p_3^*$ denote similar quantities when repressor is present. $v_0/\gamma$ in~(\ref{repr_factor}) is the stationary mean expression level at the basal transcription rate. The resulting formula for $C_R$ in terms of the stationary probabilities does not contain $v_0$ and $v_A$. { The repression factor values from the interval $0\le C_R\le 1$ cover all levels of repression strength. Negative $C_R$ values correspond to situations in which the repressor effectively becomes an activator, so we do not analyze these regimes.
	}
	
	The full expressions of $C_R$ in the nonequilibrium and equilibrium models are given in the Appendix in~\eqref{repr_factor_full_neq} and~\eqref{repr_factor_full_eq}, respectively. It follows from these expressions that the repression factor in the nonequilibrium model is positive if and only if
	\begin{equation} \label{repr-cond}
		\frac{k_3}{k_4} > \frac{k_1}{k_2},
	\end{equation}
	and it is positive for all positive values of $k_3$ and $k_4$ in the equilibrium system. If~(\ref{repr-cond}) does not hold, the presence of repressor ($c\neq 0$) shifts the balance of the probability flow in Fig.~\ref{fig:graphs}~(b) away from the chromatin closed state, i.e. the repressor acquires an activating function.
	
\begin{figure}
		\centering
		\includegraphics[width=\linewidth]{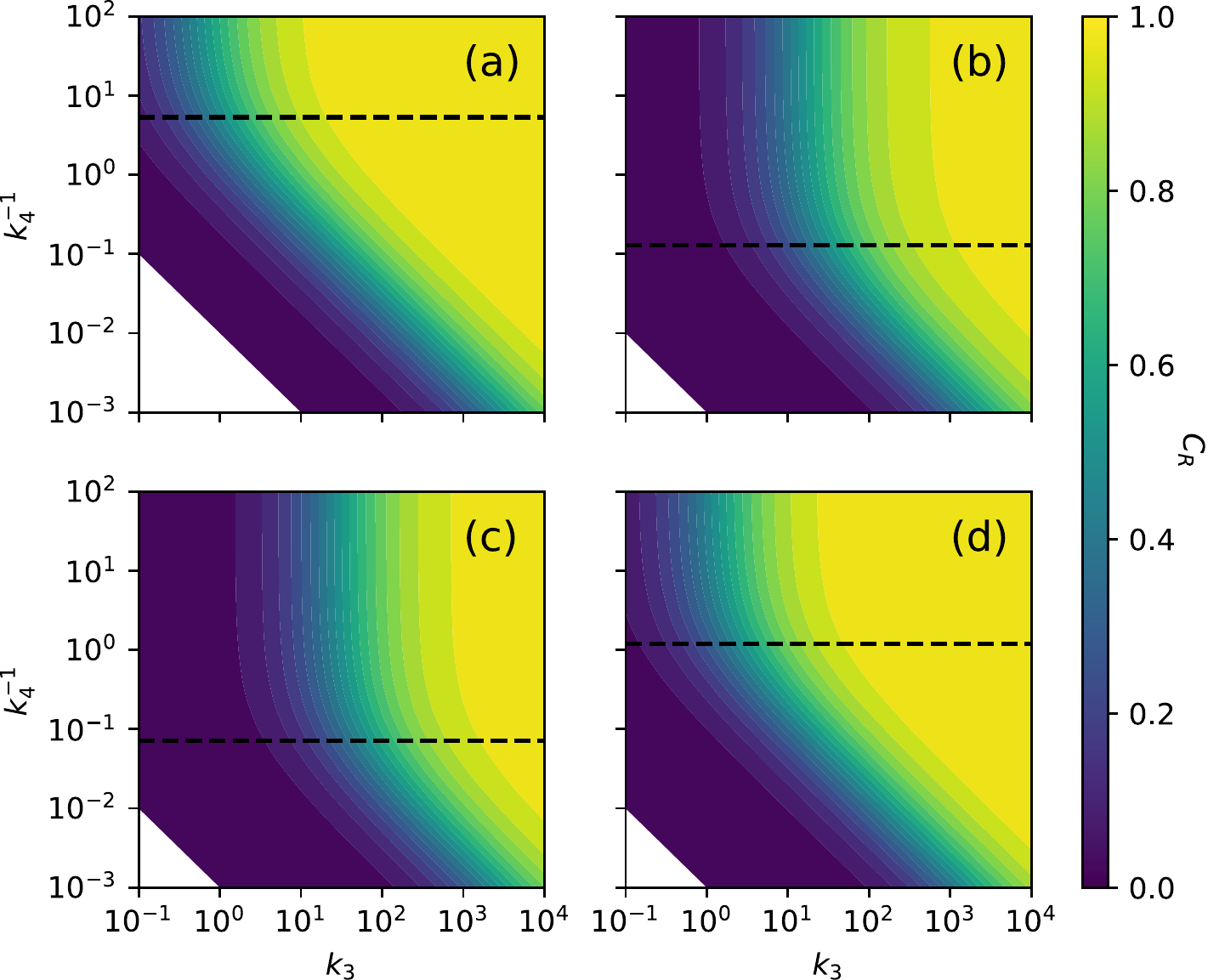}
		\caption{Dependence of the repression factor $C_R$ in the nonequilibrium model on $k_3$ and $k_4^{-1}$ for various values of other parameters. The white regions correspond to negative $C_R$. The horizontal dashed line corresponds to the critical value $k_4^*$ described in the text. {Parameters values ($a$, $b$, $c$, $d$, $k_1$, $k_2$) for each panel: (a)~(100, 50, 1000, 2, 1, 100); (b)~(100, 100, 400, 10, 1, 1000); (c)~(3000, 2000, 800, 100, 0.1, 100); (d)~(1200, 1800, 4000, 300, 0.01, 10). In all plots, $\gamma = 1$.}}
		\label{fig:repr_fact}
	\end{figure}

	Just as for the mean mRNA copy number, the repression factor in the equilibrium model depends only on the ratios of forward and backward transition rates. Hence, its dependence on $k_3$ and $k_4^{-1}$ is symmetrical, and the repression mechanisms associated with these rates cannot be distinguished at the level of mean mRNA in this model. In contrast, $C_R$ in the nonequilibrium model depends on each transition rate separately. Plotting $C_R$ as the function of $k_3$ and $k_4^{-1}$ in this model for various values of other parameters reveals different levels of repression strength at two limits, one corresponding to large values of both $k_3$ and $k_4$ and the other to small values of both rates (Fig.~\ref{fig:repr_fact}). This provides an asymmetric picture of the rate dependence in the nonequilibrium case, in which an increase in the rate of chromatin condensation ($k_3$) appears as a more efficient repression mechanism compared to a decrease in the rate of chromatin loosening ($k_4$). 
	
	We can distinguish between two different regimes for the repression factor, separated by the dashed line in Fig.~\ref{fig:repr_fact}. $C_R$ depends on $k_3$ and $k_4$ predominantly through the ratio $k_3/k_4$ at small values of $k_4^{-1}$ and is almost independent of $k_4$ at large values of $k_4^{-1}$. The critical value $k_4^*$ that approximately separates these two cases can be estimated from the analysis of $C_R$ (see~\eqref{k4_crit} in the Appendix). Values of $k_4$ in the vicinity of $k_4^*$ correspond to the cases when the repression is susceptible to both $k_3$ and $k_4$ and depends on these rates, not through their ratio.
	
	Another qualitative difference between the equilibrium and nonequilibrium models is that maximal repression in the equilibrium model can be reached by varying $k_3$ and $k_4$ alone:
	$$\lim_{k_4\to 0} C_R^{\text{eq}} = \lim_{k_3\to\infty} C_R^{\text{eq}} = 1,$$
	while this is not true for the nonequilibrium model:
	\begin{equation} \label{repr_factor_lims_neq}
		\lim_{k_4\to 0} C_R^{\text{neq}} < \lim_{k_3\to\infty} C_R^{\text{neq}} < 1,
	\end{equation} 
	and the latter limit is independent of $k_4$ (see \eqref{repr_factor_limk3} and \eqref{repr_factor_limk4} in the Appendix). Additional parameters associated with the repressor must be tuned to gain maximal repression in this model. Namely, high concentrations of repressor (large value of rate~$c$) or high specificity of the repressor binding site (small value of~$d$) are additionally required:
	\begin{eqnarray}
		\lim_{k_3,c\to\infty} C_R^{\text{neq}}\, & & = \lim_{\substack{k_3\to\infty,\\ d\to 0}} C_R^{\text{neq}} = \lim_{\substack{k_4\to 0,\\ c\to\infty}} C_R^{\text{neq}} \nonumber\\
		& & = \lim_{k_4,d\to 0} C_R^{\text{neq}} = 1.
	\end{eqnarray}
	
	In order to understand how the repression mechanisms associated with $k_3$ and $k_4$ influence the noise in gene expression, we analyzed the Fano factor, or the variance-to-mean ratio, as a function of these rates. 
	The full analytical expression of the Fano factor is cumbersome and is not given here. An asymptotic analysis reveals that increasing the rate $k_3$ yields smaller Fano factor values than decreasing $k_4$ in both models:
	\begin{equation} \label{fano_lims}
		\lim_{k_4 \to 0}\frac{\sigma^2}{\mu} > \lim_{k_3 \to \infty}\frac{\sigma^2}{\mu},
	\end{equation}
	and the latter limit is independent of $k_4$. This is an indication that the two repression mechanisms contribute to gene expression noise asymmetrically in both equilibrium and nonequilibrium models.
	
\begin{figure}
		\centering
		\includegraphics[width=\linewidth]{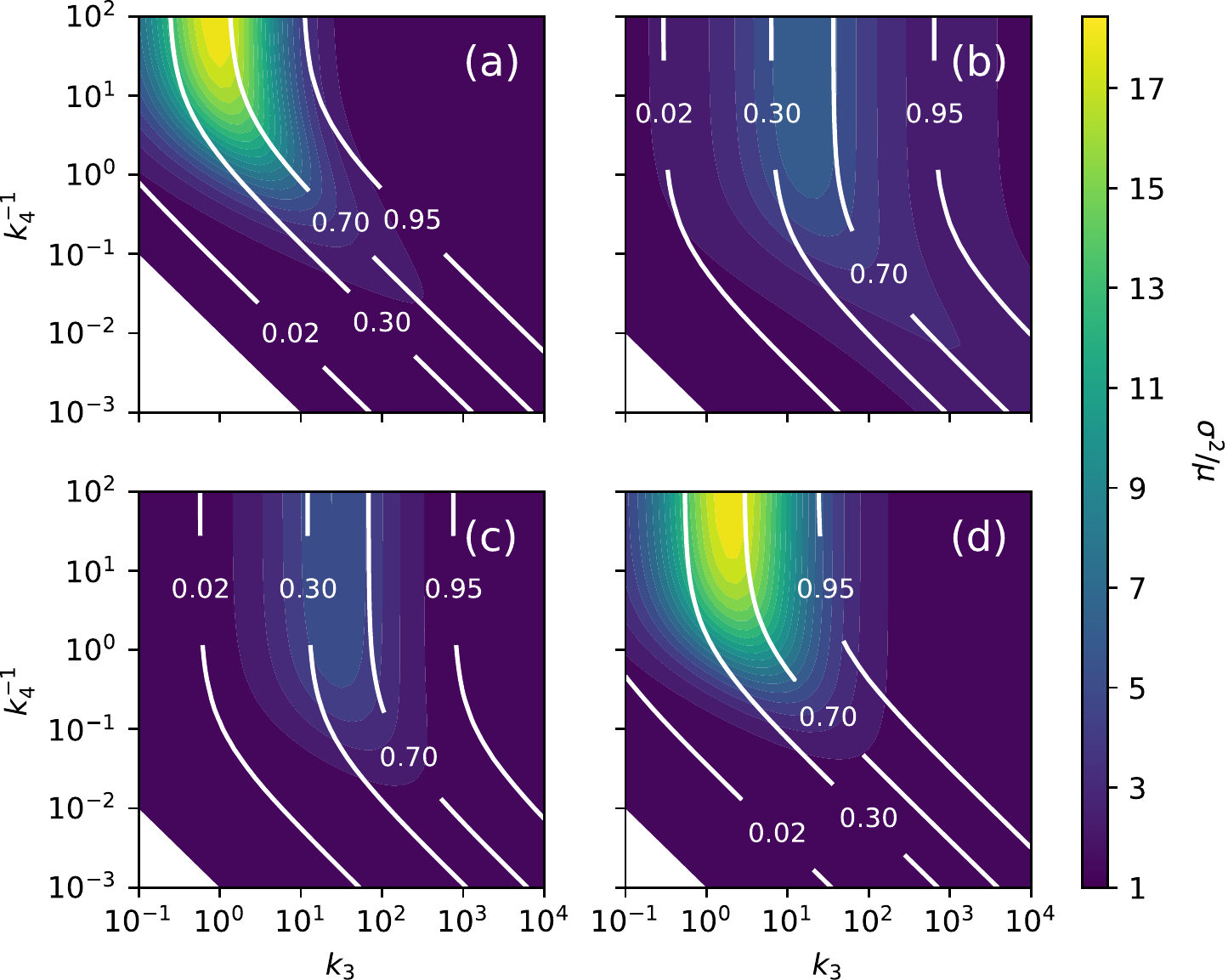}
		\caption{Dependence of the Fano factor on $k_3$ and $k_4^{-1}$ for various values of other parameters in the nonequilibrium model. The white lines show levels of constant repression factor $C_R$ (and, therefore, constant mean expression). The white regions correspond to negative $C_R$. { Parameter values for each panel consist of the values from Fig.~\ref{fig:repr_fact} and the following values of ($v_0$, $v_A$): (a)~(10, 100); (b)~(10, 300); (c)~(20, 400); (d)~(0, 200).} In all plots, $\gamma = 1$.}
		\label{fig:fano_6neq}
	\end{figure}

\begin{figure}
		\centering
		\includegraphics[width=\linewidth]{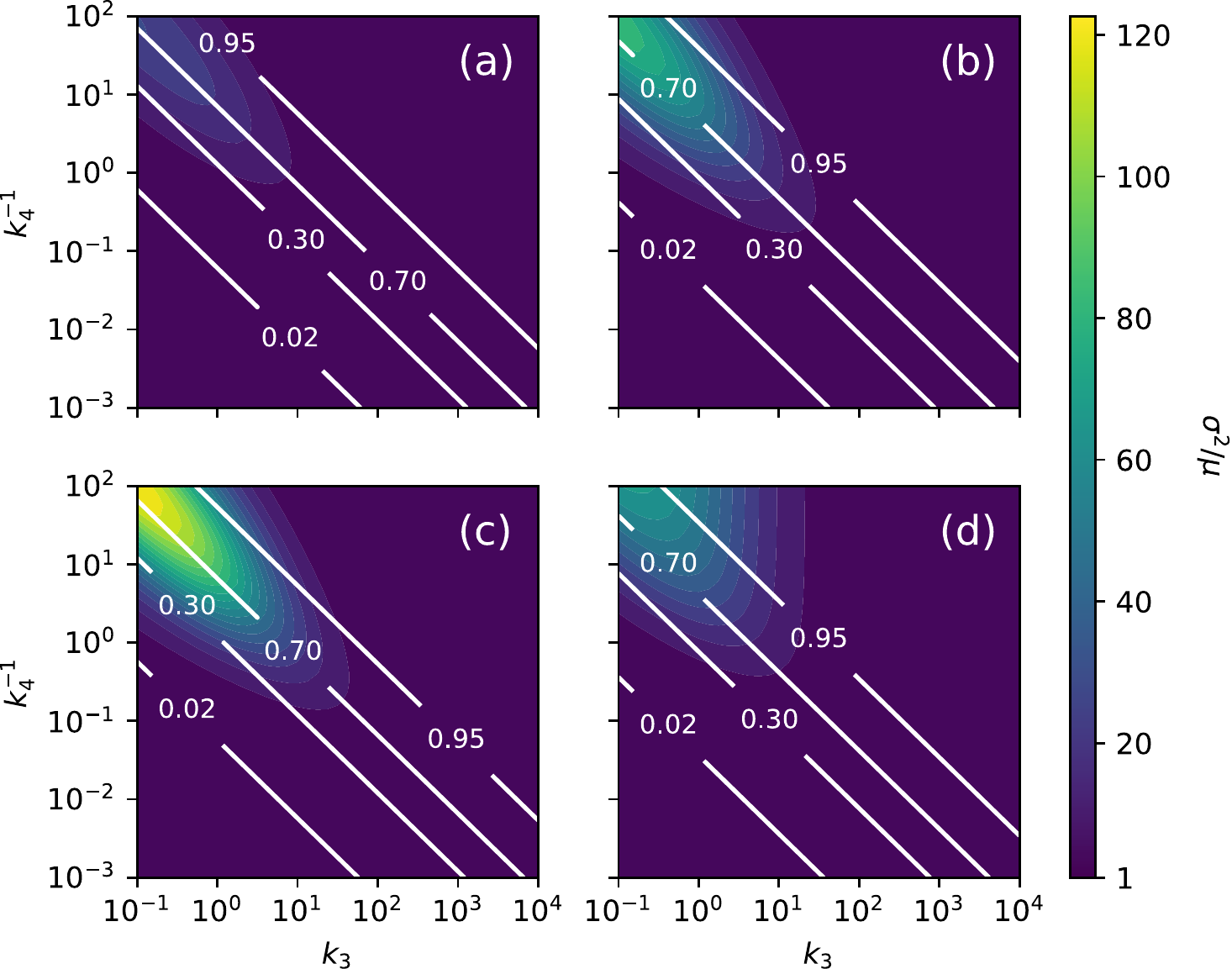}
		\caption{The same as in Fig.~\ref{fig:fano_6neq} but for the equilibrium model. {Parameter values for each panel are as in Figures~\ref{fig:repr_fact} and~\ref{fig:fano_6neq} except that $k_1$ and $k_2$ should be omitted.}}
		\label{fig:fano_5eq}
	\end{figure}

	The simulation of the Fano factor dependence on $k_3$ and $k_4^{-1}$ for various values of model parameters gives a visual representation of this asymmetry (Figs.~\ref{fig:fano_6neq} and~\ref{fig:fano_5eq}). The figures show that repression due to a decrease in the probability of chromatin loosening (large $k_4^{-1}$) leads to much greater noise than repression due to an increase in the probability of chromatin condensation (large $k_3$), even at the same repression factor levels. 
	
\begin{figure}
		\centering
		\includegraphics[width=.95\linewidth]{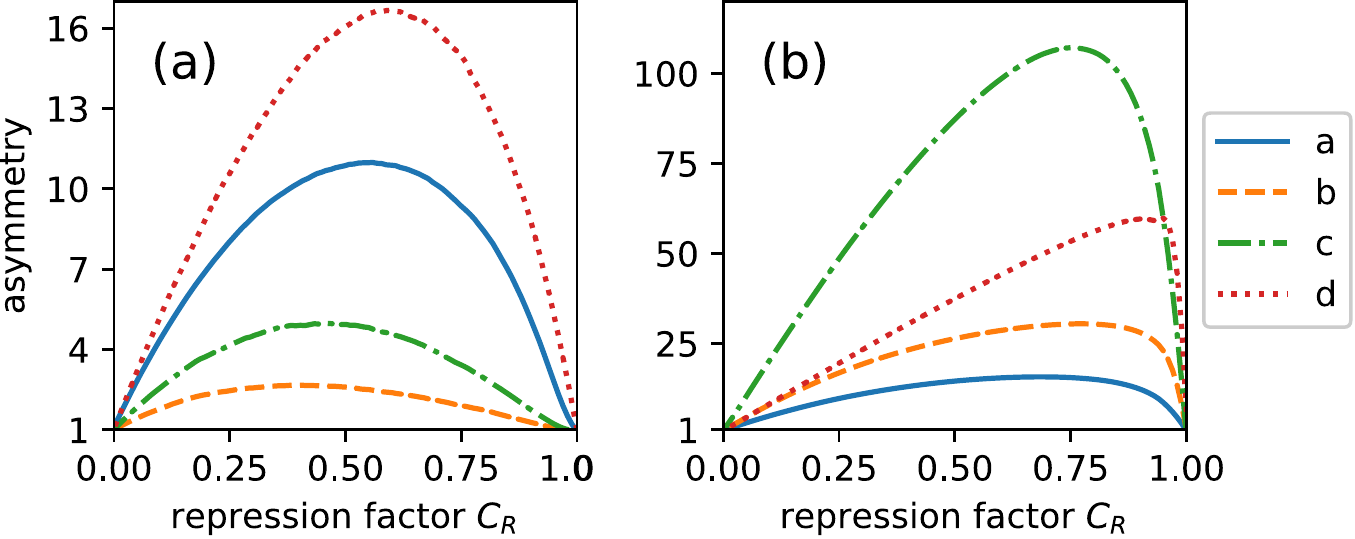}
		\caption{The asymmetry of Fano factor as a function of the repression factor {for (a) the nonequilibrium model  and (b) the equilibrium model}.
			{ The curves correspond to parameter values taken from the captions of Figures~\ref{fig:fano_6neq} (nonequilibrium model) and~\ref{fig:fano_5eq} (equilibrium model) as indicated in the inset.}}
		\label{fig:asym}
	\end{figure}

	We quantified the asymmetry of the Fano factor dependence on $k_3$ and $k_4^{-1}$ as the ratio of its maximal to its minimal value along the fixed repression factor level lines and plotted this ratio as a function of the repression factor (Fig.~\ref{fig:asym}). 
	The two models exhibit qualitatively different distributions of this asymmetry over the repression levels. 
	The asymmetry in the nonequilibrium model has a maximum at moderate repression levels (around $C_R=0.5$), while in the equilibrium model this maximum is shifted to the large repression levels, and, in the case of $v_0=0$, this maximum is almost at $C_R=1$ (curve d in Fig.~\ref{fig:asym}~(b)).

	\section{Application to gene expression data}
	We further analyze the difference between the equilibrium and nonequilibrium models by applying the models to previously published data on gene expression in an experimental setup that is very close to the system from Fig.~\ref{fig:graphs}~(a)~\cite{Fakhouri2010}. We consider five genetic constructs from that experiment (constructs 1--5 from Figure 2 in Ref.~\cite{Fakhouri2010}) regulating the expression of the reporter gene \textit{lacZ} in transgenic \textit{Drosophila} lines.
	Each construct consists of a pair of sites binding the short-range repressor Giant, a pair of sites binding the activating TF Twist, and a pair of sites binding the activating TF Dorsal. The activator sites are located next to each other, and the same is true for the repressor sites. The constructs differ from each other by the distance between the group of repressor sites and the group of activator sites, which takes the following values in constructs 1 through 5: 0, 25, 35, 50, and 60 base pairs. The increasing distance between the activator and repressor sites is associated with the reducing repression strength of Giant. Following the authors of ref.~\cite{Fakhouri2010}, we assume that the two consecutive Giant binding sites are a single repressor site and the four consecutive sites of the activating TFs are a single activator site. Each construct corresponds to the system depicted in Fig.~\ref{fig:graphs}~(a) under this assumption, so we can apply the modeling formalism described above.
	
	The expression data for each construct contained the normalized \textit{lacZ} expression values for a set of normalized Giant concentrations~\cite{Fakhouri2010}. We preprocessed these data obtaining estimates of the absolute \textit{lacZ} expression levels in the mRNA copy numbers, as described in the Appendix. Next, we split the processed expression values into 20 bins corresponding to different Giant concentrations, and we calculated the mean and variance within each bin, separately for each construct.
	We fitted the analytically derived mean and variance from~(\ref{mrna_mean_steady}) and~(\ref{mrna_variance_steady}) to these experimental mean and variance values.

	\subsection{Comparison of models}
	
	During the model fitting, we optimized parameter values by minimizing the root-mean-square error (RMSE) as an objective function:
	\begin{equation}\label{rmse}
		\text{RMSE} = \sqrt{\frac{1}{2n} \sum_{i=1}^n \left( \left(\mu_i - \tilde\mu_i\right)^2 + \left(\sigma_i - \tilde\sigma_i\right)^2 \right)},
	\end{equation}
	where $\tilde\mu_i$ and $\tilde\sigma_i$ denote the experimental mean and standard deviation values for the $i$th bin ($i$th value of Giant concentration), $\mu_i$ and $\sigma_i$ are the model predictions of these quantities from~(\ref{mrna_mean_steady}) and~(\ref{mrna_variance_steady}), and $n$ is the total number of bins. We discarded several bins corresponding to large Giant concentrations from the data for constructs 4 and 5, as the expression variance demonstrated an anomalous jump in these bins, which we interpret either as an artifact or as associated with an unknown regulator (see discussion in the Appendix). 
	
	The experimentally measured values of Giant concentration are used as the [R] concentration in~(\ref{a_c}), with $c_0$ as a free parameter. As Twist and Dorsal are ubiquitously expressed transcription factors, we do not estimate the activator concentration [A] explicitly and leave $a$ as a free parameter~\cite{Fakhouri2010}.
	In order to reduce the number of free parameters, we set the basal rate of mRNA production to zero ($v_0 = 0$).
	Therefore, the full list of free parameters consists of transition rates ($a,\ b,\ c_0,\ d,\ k_3$, and $k_4$ in both models and additionally $k_1$ and $k_2$ in the nonequilibrium model), transcription rate $v_A$, and mRNA degradation rate $\gamma$. As the formulas for the mean and variance will not change if all parameters are divided by $\gamma$, we set $\gamma = 1$ and interpret other parameters as divided by $\gamma$. We assume that rates $k_3$ and $k_4$, which define the efficacy of the repressor, are specific for each construct while all other model parameters are the same for all constructs.
	
\begin{figure*}
		\centering
		\includegraphics[width=.9\linewidth]{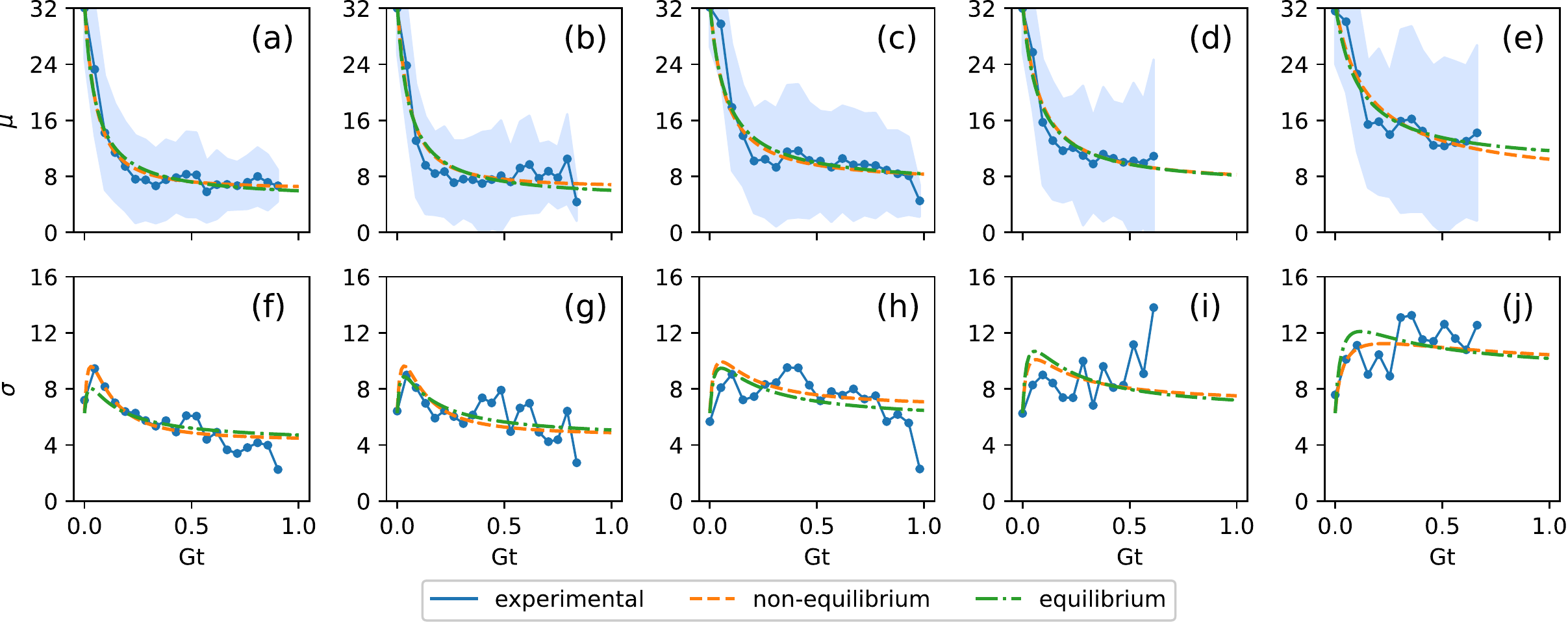}
		\caption{{Mean expression and standard deviation in the} best solutions found by numerical optimization in the nonequilibrium and equilibrium models in comparison with data. Gt, normalized Giant concentration. The filled area in the first row of panels shows the $\mu\pm \sigma$ interval. { Columns (a, f)--(e, j) correspond to constructs one through five, respectively, in the data.}}
		\label{fig:6neq_5eq_mean_var}
	\end{figure*}

	We performed 200 optimization runs for each model using the dual annealing algorithm from SciPy library~\cite{2020SciPy-NMeth, Xiang1997}. The best solutions in both models show a good correspondence to the data (Fig.~\ref{fig:6neq_5eq_mean_var}).
	The two solutions are visually close to each other, but the nonequilibrium one demonstrates slightly smaller errors for most constructs in the data and for most Gt concentrations (Fig.~S1, Supplemental Material~\bibnote[SupplMat]{See Supplemental Material at [URL will be inserted by publisher] for additional figures and extended models description}). Another qualitative difference between the models is that the nonequilibrium model effectively samples a wider range of Fano factor asymmetry values in parameter optimization (Fig.~S2~\cite{SupplMat}).
	
\begin{figure}
		\centering
		\includegraphics[width=\linewidth]{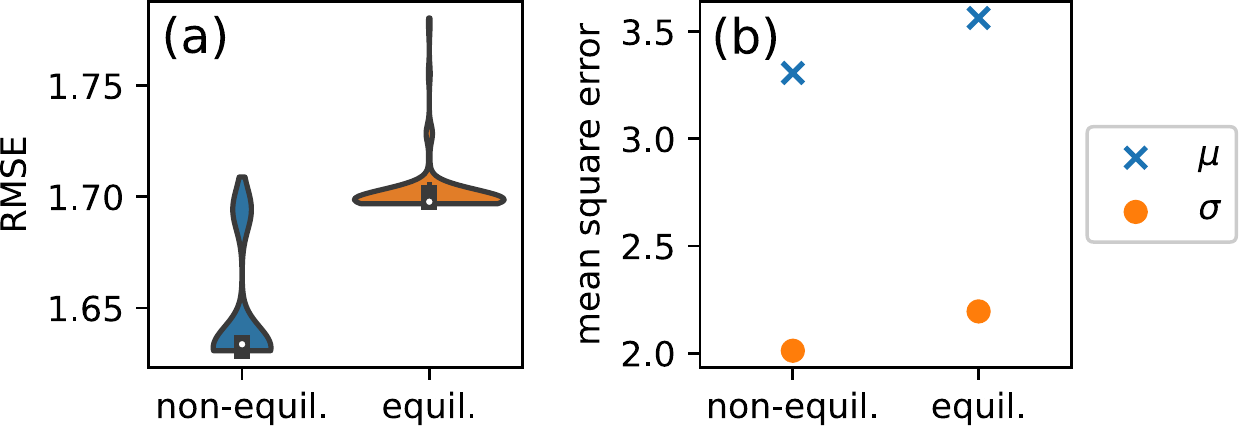}
		\caption{{ Violin plots} for RMSE values (a) and mean square error of the mean and standard deviation separately (b) obtained in 200 optimization runs in the equilibrium and nonequilibrium models.}
		\label{fig:6neq_5eq_error}
	\end{figure}

	To compare the models quantitatively, but reduce possible overfitting, we used an ensemble approach and analyzed all optimization results simultaneously, not only the best one. In the majority of optimization runs, the nonequilibrium model resulted in smaller RMSE values than the minimal RMSE obtained in the equilibrium model, and this holds both for the mean and for the variance part of the objective function separately (Fig.~\ref{fig:6neq_5eq_error}). The optimization for the nonequilibrium model yielded RMSE = 1.63 in most cases (154 runs out of 200), while the lowest RMSE value obtained for the equilibrium model was 1.7. 
	
	As the nonequilibrium model has two additional parameters ($k_1$ and $k_2$) compared to the equilibrium one, we scored the performance of the two models using the Akaike information criterion with correction for small data samples~\cite{Burnham2002}:
	\begin{equation}\label{aic}
		\text{AICc} = 2k + 2n \ln(\text{RMSE}^2_{\text{min}}) + \frac{2k(k+1)}{2n - k - 1},
	\end{equation}
	where $k$ is the number of model parameters and $2n$ is the total number of data points. AICc scores for the nonequilibrium and equilibrium models are equal to 191.2 and 200.8, respectively. 
	
	Fig.~\ref{fig:6neq_5eq_mean_var} shows that the standard deviation fit is not satisfactory for construct 4 at moderate concentrations of Giant. To ensure that this flaw does not affect our conclusions, we performed additional fits using data without construct 4. We obtained similar parameter values and confirmed the advantage of the nonequilibrium model (Fig.~S3~\cite{SupplMat}).
	
	We got similar results when the models were fitted only by their mean values, i.e. when noise was discarded in the data. In this case, the objective function is obtained from~(\ref{rmse}) by removing the difference between standard deviations.
	The nonequilibrium model has 5 parameters more in this setting, because the equilibrium model depends only on the ratios of the forward and backward transition rates at the level of the mean. However, the AICc score for the nonequilibrium model (74.6) is significantly lower than the score for the equilibrium model (117.4). { We additionally fitted using the functional that contains the inverse variance as weights, as follows:
		\begin{equation}
			\text{RMSE} = \sqrt{\frac{1}{n} \sum_{i=1}^n \frac{\left(\mu_i - \tilde\mu_i\right)^2}{\tilde\sigma_i^2}}.
		\end{equation}
		This computational experiment also resulted in} better performance of the nonequilibrium model compared to the equilibrium one in terms of the AICc score ($-314$ vs. $-287$, respectively).
	
	{ 
		The nonequilibrium model reduces the RMSE score by 4\% compared to the equilibrium case (Fig.~\ref{fig:6neq_5eq_error}), which can be estimated as a relatively small value given the noisy gene expression. However, as we show in section~\ref{sec:mechanisms}, the main advantage of the nonequilibrium approach is its ability to distinguish between different mechanisms of repression.
		
	}


	\subsection{Entropy production rate}
	
	The gene state transition graph can be related to the thermodynamic quantities by introducing 
	the probability flux $F_{ij}$ from $i$-th to $j$-th state and corresponding thermodynamic force $A_{ij}$~\cite{Qian2016}:
	\begin{equation}\label{flux_force}
		F_{ij} = p_i r_{i\to j} - p_j r_{j\to i}, \quad A_{ij} = \ln{\frac{p_i r_{i\to j}}{p_j r_{j\to i}}},
	\end{equation}
	where, as before, $p_i$ denotes the probability of the $i$-th state, and $r_{i\to j}$ is the rate of transition from $i$-th to $j$-th state. The detailed balance requires $F_{ij} = A_{ij} = 0$ for all $i$ and $j$.
	Using these variables, the internal entropy production rate can be calculated as follows:
	\begin{equation}\label{entropy_prod}
		\frac{d_i S}{dt} = \frac12 k_B \sum_{i,j} F_{ij}A_{ij},
	\end{equation}
	where $S$ is the entropy, and $k_B$ is the Boltzmann constant. This expression can be rewritten in terms of the fundamental set of the graph cycles emphasizing the key role of cycles in the entropy production~\cite{Schnakenberg1976,Jiang2004}.
	An increase in entropy indicates irreversible processes that dissipate energy. The dissipation power can then be estimated by multiplying the internal entropy production rate by the ambient temperature.
	
\begin{figure}[h]
		\includegraphics[width=.75\linewidth]{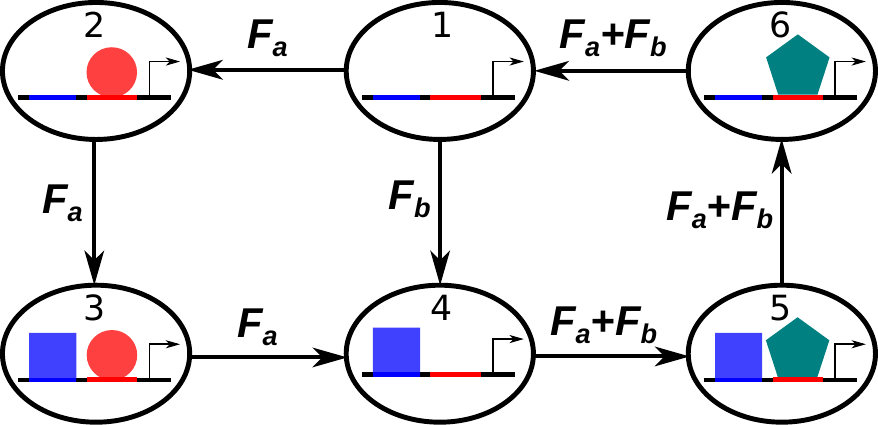}
		\caption{{ Probability fluxes $F_{ij}$ in the nonequilibrium model for parameter values corresponding to the best fit and the first construct in the data; $F_a = 3.2 \cdot 10^{-4}$, $F_b = 3.6 \cdot 10^{-4}$. These values were calculated using $\gamma=0.4\cdot 10^{-2}$ sec.$^{-1}$ (see Appendix, section~\ref{sec:param_ident}).}}
		\label{fig:fluxes}
	\end{figure}

	To see how the detailed balance is broken in the nonequilibrium model, we visualized the probability fluxes $F_{ij}$ in Fig.~\ref{fig:fluxes} for the parameter values corresponding to the first construct in the data. The figure shows that the flux is non-zero between each pair of states in the state transition graph, demonstrating that the gene operates in a nonequilibrium regime. 
	{ The total dissipation power of the system from Fig.~\ref{fig:fluxes} is $2.3\cdot 10^{-23}$ W at temperature $T=298$~K, which approximately corresponds to the standard free energy of one ATP hydrolysis reaction ($\sim5\cdot 10^{-20}$~J~\cite{Rosing1972}) expended each 2000 seconds. The small value obtained in this estimate depends on the chosen data normalization method and other approximations made in our study, so it should be treated with caution. We can expect a significant increase in the dissipation power in more realistic regulatory modules consisting of many sites for many transcription factors. Similar results for parameter values corresponding to other constructs are shown in Figs.~S13 and S14~\cite{SupplMat}.}

	\subsection{Distinguishing between short-range repression mechanisms}\label{sec:mechanisms}
	
	We showed above that the short-range repression mechanisms associated with the rates $k_3$ and $k_4$ can be separated in the nonequilibrium model both at the level of mean expression and its variance, and we demonstrated that the repression strength depends on these constants asymmetrically. In this section, we use the nonequilibrium model and expression data to identify which of these mechanisms is a primary source of repression in the biological system. The first mechanism is present if the rate values inferred from the data obey the inequality $k_3>k_1$, which means that the bound repressor facilitates the recruiting of factors associated with chromatin condensation. The second mechanism is present if $k_4<k_2$, which means that the bound repressor hampers the local loosening of chromatin.
	
\begin{figure}
		\centering
		\includegraphics[width=.65\linewidth]{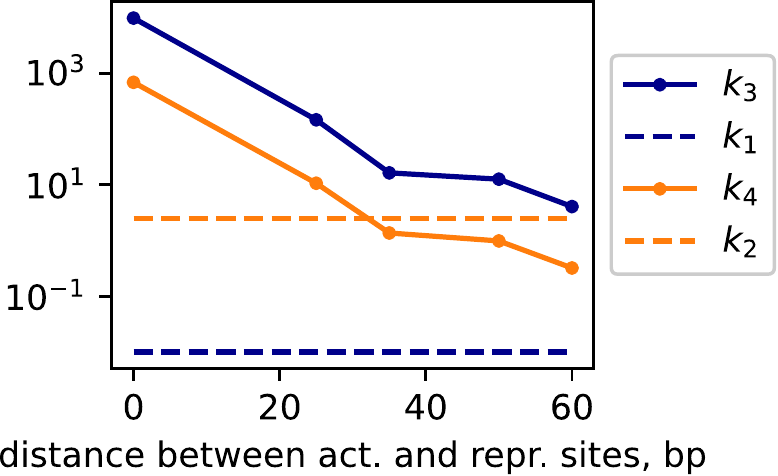}
		\caption{Dependence of the construct-specific rates $k_3$ and $k_4$ on distance between the activator site and repressor site as a result of parameter optimization in the nonequilibrium model. The dashed horizontal lines indicate values of the construct-independent rates $k_1$ and $k_2$.}
		\label{fig:k3_k4_vs_dist}
	\end{figure}

	A typical relation between the rates as found from optimization is shown in Fig.~\ref{fig:k3_k4_vs_dist} for all constructs, i.e. for all distances between the activator and repressor sites. The pattern is that $k_3>k_1$ and $k_3$ decreases monotonically with an increase in distance between the binding sites. This means that the first mechanism is present in the system, and a larger distance expectedly corresponds to the lower impact of this mechanism on the total repression. On the other hand, the decrease of $k_4$ with distance is counterintuitive, since a more distantly bound repressor should be less effective in holding the condensed chromatin state on the activator site, thus leading to larger $k_4$. Moreover, $k_4 > k_2$ for small distances, i.e. the condensed state becomes less stable in the presence of a repressor. 
	
	A possible explanation for this behavior of $k_4$ can be related to a hypothesis that, in addition to providing the fixation of the condensed chromatin state as a repression mechanism, a repressor located too close to the activator site also can destabilize this state by an independent mechanism. In this case, the distance dependence of $k_4$ and its alternating relation to $k_2$ in Fig.~\ref{fig:k3_k4_vs_dist} can be an emergent property resulting from the two counteracting mechanisms working simultaneously. However, we believe that another explanation is more likely. We observed a high correlation between values of $k_3$ and $k_4$ found in multiple optimization runs for a single construct {(Figure S7~\cite{SupplMat})}, which can be attributed to a certain level of practical non-identifiability of these parameters (see the Appendix for more details). This correlation can have the same nature as the correlation between $k_3$ and $k_4$ values for all the constructs shown in Fig.~\ref{fig:k3_k4_vs_dist}. As $k_3$ behaves more expectedly, we may suggest that $k_3$ serves as a more effective repression parameter in the model, while $k_4$ just follows the values of the former as a less identifiable parameter.
	
	Plotting the best fit values of $k_3$ and $k_4$ in an analog of Fig.~\ref{fig:repr_fact} shows that the $k_4$ values are in the vicinity of $k_4^*$ for several constructs (Fig.~S4~\cite{SupplMat}). This means that the observed correlation between $k_3$ and $k_4$ in the fit cannot be explained by the dependence of the repression factor on $k_3$ and $k_4$ through their ratio. The parameter values found by optimization for most constructs lie in a region where $k_3$ and $k_4$ are effectively uncoupled in the repression factor.
	
\begin{figure}
		\centering
		\includegraphics[width=.97\linewidth]{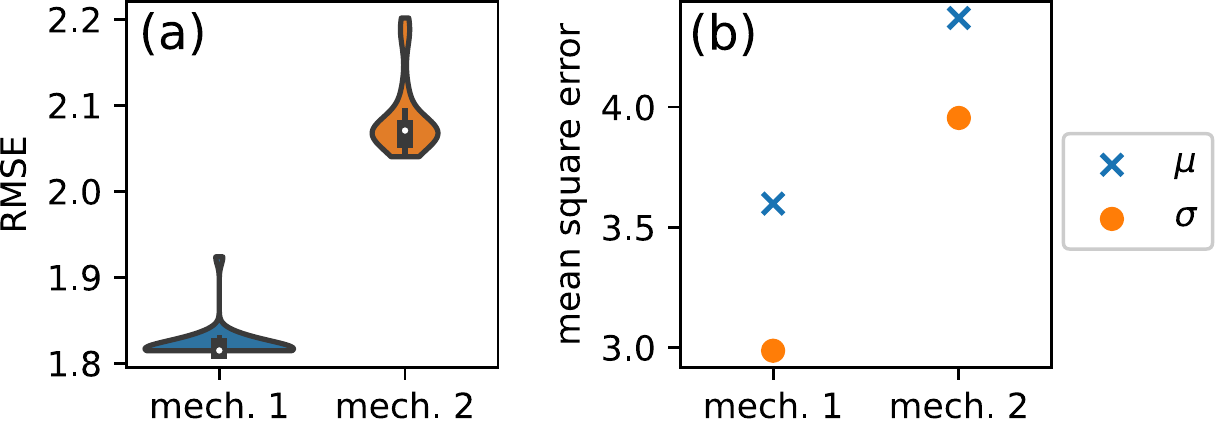}
		\caption{{ Violin plots} for RMSE values (a) and mean square error of the mean and standard deviation (b) obtained in a set of 100 optimization runs in the nonequilibrium model with fixed $k_4$ (mechanism 1) and $k_3$ (mechanism 2).}
		\label{fig:mech1_mech2_error}
	\end{figure}

	In order to identify the mechanism primarily responsible for repression and to get rid of possible non-identifiability issues, we performed parameter optimization in the nonequilibrium model with either $k_3$ or $k_4$ fixed across the constructs. In the first computational experiment, we set $k_4 = k_2$ and left $k_3$ to be construct specific. This setting means that only the first repression mechanism is present in the system. In the second experiment, we set $k_3 = k_1$ and left $k_4$ to be construct-specific, so that repression in the system was only due to the second mechanism. The optimization results showed that the model with the first repression mechanism as the sole source of repression is associated with a significantly smaller RMSE than the model with the second mechanism (Fig.~\ref{fig:mech1_mech2_error}). This suggests that repression by increasing $k_3$ is more effective in the nonequilibrium model for gene regulation by Giant.
	
	\subsection{Extended models}
	We investigated extended models implementing a modified version of the state transition graph from Fig.~\ref{fig:graphs}~(b), in which the repressor binds its site in the presence of nucleosome (transition from state 6 to state 5) with the rate $c_1$ ($c_1\neq c$) and unbinds (transition from state 5 to state 6) with the rate $d_1$ ($d_1\neq d$). This extension relies on a reasonable assumption that the DNA compaction state alters the DNA binding and unbinding kinetics of the repressor. We also considered a possibility that the basal transcription rate $v_0$ is not zero, and we added this rate as an additional parameter to the extended models. The modified transition graph can provide repression under equilibrium conditions, so we keep all six states of the graph in the extended equilibrium model. 
	
	We optimized the parameters in the extended models in the same way as before (Section II of Supplemental Material~\cite{SupplMat}). The nonequilibrium model demonstrated better performance on the expression data than the equilibrium one, both in terms of the total error (Fig.~S10~\cite{SupplMat}) and AICc score, showing an essentially better quantitative correspondence with the data in the mean expression, but a slightly worse correspondence in the expression variance (Figs.~S10 and~S11~\cite{SupplMat}). We simulated the two repression mechanisms by setting either $k_3=k_1$ or $k_4=k_2$ in the extended nonequilibrium model, and we performed the same computational experiment in the extended equilibrium model, since the new equilibrium model contains the same parameters as the nonequilibrium one and allows $k_i$ separation in the variance. The first mechanism was associated with a smaller error than the second one in the nonequilibrium model, and the errors in the equilibrium model were almost the same for both mechanisms and were close to the error in the nonequilibrium model associated with the second mechanism (Fig.~S16~\cite{SupplMat}). Therefore, the mechanisms are indistinguishable in the extended equilibrium model, and the first repression mechanism is more preferable according to the nonequilibrium approach.
	
	\section{Discussion}
	Regulation of gene expression involves energy-dissipating processes, therefore it operates away from equilibrium~\cite{Gunawardena2020}. However, quantitative models based on the assumption that some of these processes stay in a thermodynamic equilibrium showed tremendous success in describing spatio-temporal expression patterns of many genes in many organisms~\cite{Bintu2005, Bintu2005a, Segal2008, Martinez2014, Samee2014, Kozlov2015, Hoermann2016, Bertolino2016}. The equilibrium assumption brings simplicity in terms of a reduced number of free parameters into the gene expression models, which already incorporate quite a few details about gene regulation. The costs for this simplification can only be determined by examining both equilibrium and nonequilibrium models in a truly comparative study.
	
	We formulated a simple model of gene regulation involving short-range repression, and we compared the equilibrium and nonequilibrium representations of this model. We showed that, under the choice of parameters as in Fig.~\ref{fig:graphs}, the full graph including all possible states of the regulatory DNA region is inherently nonequilibrium, because the introduction of detailed balance into the graph leads to effectively no repression in the system. This forces the use of a reduced state transition graph within the equilibrium framework, leaving one regulatory state aside. Therefore, the use of the equilibrium approximation for this system 
	implies qualitative changes in the full picture of gene transcription, and not only a smaller number of parameters.
	Alternatively, one can apply the extended equilibrium model, which has all six states in the transition graph, but this model requires a larger number of free parameters.
	
	As a short-range repressor provides local chromatin compaction, preventing the activator from binding DNA or from recruiting additional factors that bind to the basal promoter, two mechanisms can be distinguished for how this compaction is achieved~\cite{Li2011}. The repressor may recruit factors mediating chromatin condensation or block proteins responsible for chromatin loosening. In the model, these two mechanisms can be formalized with an increase in the rate $k_3$ or a decrease of the rate $k_4$, respectively. The equilibrium model does not separate these mechanisms at the level of the expression mean.
	In contrast, we showed that the repression efficiency in the nonequilibrium model depends on $k_3$ and $k_4^{-1}$ asymmetrically, which means that only this formalism can be used to infer a possible difference between the repression mechanisms. On the other hand, both models are applicable in analyzing the influence of the different repression mechanisms on the expression noise, and they lead to the conclusion that the $k_3$-related mechanism provides less noisy expression.
	
	The qualitative difference in how the repression mechanisms are implemented in the models also appears in the fact that the maximal repression in the equilibrium model can be achieved at either infinitely large $k_3$ or infinitely small $k_4$, irrespective of the values of all other parameters. The nonequilibrium model demands the binding and unbinding rates for the repressor to be additionally tuned to gain the maximal repression. Thus, the interaction between the repressor and its binding site on the DNA comprises an independent source for repression in the nonequilibrium model, which cannot be compensated by any action of a bound repressor, in contrast to the equilibrium formalism.
	
	We showed that the nonequilibrium model better describes gene expression data in the context of regulation by the Giant transcription factor in \textit{Drosophila} development. This model is more accurate than the equilibrium one in estimating both the mean expression and expression variance of the target gene. As the total number of free parameters in both models is relatively large, our estimates of parameter values demonstrate some variation. An efficient way to increase the confidence in examining various hypotheses in this situation is to use an ensemble approach, taking into account all possible combinations of parameter values and corresponding model performance scores in the analysis~\cite{Samee2015}. Moreover, most of the parameter values found by parameter optimization yield very close performance scores. We also reduced parameter variation by fixing some parameters without affecting the quality of fitting (see the Appendix).
	
	Despite the fact that the quantitative measures indicate outperformance of the nonequilibrium model over the equilibrium one on the expression data, the solutions in the two models are visually close. Therefore, the new model should not be considered as a better descriptive tool, at least for the chosen data, especially taking into account the larger number of parameters. However, the main advantage is that the nonequilibrium model brings a possibility to verify biological hypotheses that can be impossible to examine using the equilibrium framework. We showed that the distinction between two alternative mechanisms of repression on the expression data is indeed possible in the nonequilibrium model, but not in the equilibrium one.
	
	The simulation of the two repression mechanisms on the expression data reveals that the rate $k_3$ is more preferable as a repression parameter than $k_4$. Therefore, the modeling predicts that the short-range repression by Giant is most probably associated with the recruiting of factors mediating chromatin condensation, rather than with the blocking of proteins responsible for chromatin loosening. The preference of this mechanism can theoretically be related to two analytical results shown in our study. First, repression by increasing $k_3$ is stronger (in terms of the repression factor) than by decreasing $k_4$ in a wide range of values of other parameters. Second, the $k_3$-related repression provides lower expression noise. Giant belongs to the TF family that regulates the segmentation of the \textit{Drosophila} embryo as part of the embryonic developmental program, which is a process involving highly coordinated expression patterns of many genes. These specifics may require the selection of more efficient and precise regulatory mechanisms in the course of evolution. 
	
	Violation of the detailed balance in the fitted nonequilibrium model entails the internal production of entropy in time, which in theory means the dissipation of energy in the system. This energy flux is needed to maintain the system in a stationary state. The nonequilibrium model predicts a specific pattern of the probability flux in the gene state transition graph. Experimental estimates of the energy expenditure associated with gene regulation can help validate these model predictions or can be used as additional data for model calibration.
	
	We used the nonequilibrium formalism to study short-range repression in the context of a regulatory region with a fairly simple architecture. This approach can straightforwardly be generalized to more complex regulatory modules, consisting of multiple activating and repressing sites and/or involving cooperative interactions between TFs. The rise in complexity leads to disadvantages of the increasing number of parameters in the model. This problem can partially be solved by decomposing a complex interaction graph into more or less independent modules so that each module can be associated with a model of a moderate complexity~\cite{Ahsendorf2014}. For example, a similar categorization approach has proven effective in finding basic building blocks that govern the logic of how the TF--DNA molecular configurations are formed for complex promoters~\cite{Ezer2014}. Another way to generalize our model is to estimate the gene state transition rates as functions of microscopic parameters which would accommodate processes associated with possible energy dissipation.
	
	It is also important to improve our modeling results by distinguishing between the intrinsic and extrinsic sources of noise in the data. The models considered in this paper assume that all observed variability is entirely intrinsic, i.e. it stems from the stochastic nature of gene transcription and regulation. Making this distinction may require more sophisticated models as well as more advanced data, since the in situ data we used for model validation have limitations in estimating true concentration levels and their variability. However, we confirmed the conclusions on the separation of short-range repression mechanisms in the models fitted only to the mean expression, so our results are reliable even after considering the possible inaccuracy in treating noise in the data.

	\section{Appendix}
	
	\subsection{Long formulas}
	Here, we show the components $\rho_i^*$ of the Laplacian matrix kernel from eq.~\eqref{laplace_ker} normalized by $\rho_1^*$. For the equilibrium model:
	\begin{equation} \label{stat_weights_eq}
		\rho^*_1 = 1,\ \rho^*_2 = \frac{a}{b},\ \rho^*_3 = \frac{ac}{bd},\ \rho^*_4 = \frac{c}{d},\ \rho^*_5 = \frac{ck_3}{dk_4}.
	\end{equation}
	For the nonequilibrium model:
	\begin{eqnarray}
		&&\rho^*_1 = 1,\quad \rho^*_2 = \frac{a}{b}\left(1 - \frac{c\alpha}{F}\right),\nonumber\\
		&&\rho^*_3 = \frac{ac}{bd}\left(1 - \frac{(b+c)\alpha}{F}\right),\nonumber\\
		&&\rho^*_4 = \frac{c}{d}\left(1 - \frac{(b+c+d)\alpha}{F}\right),\nonumber\\ 
		&&\rho^*_5 = \frac{ck_3}{dk_4}\left(1 - \frac{\left(b+c+d + s\frac{d}{k_3}\right)\alpha}{F}\right),\nonumber\\
		&&\rho^*_6 = \frac{k_3}{k_4}\left(1 - \frac{\left(b+c+d + s\frac{k_4+d}{k_3}\right)\alpha}{F}\right), \label{stat_weights_neq}
	\end{eqnarray}
	where $s = a+b+c+d$, $\alpha = k_3 k_2 - k_4 k_1$, and ${F = s(dk_2+ck_4+k_2k_4) + (b+c+d)k_2k_3}$. The probabilities of gene states are then $p_i^* = \rho_i^* / \sum \rho_i^*$. Note that the detailed balance holds if $\alpha=0$.
	
	The repression factor in the nonequilibrium and equilibrium models takes the form:
	\begin{eqnarray} 
		\label{repr_factor_full_neq}
		C_R^{\text{neq}} &=& \frac{\frac{c}{d}\left(\frac{k_3}{k_4} - \frac{k_1}{k_2}\right)k_4 [s(c+d+k_2) + uk_1]}{F \sum_{i=1}^6 \rho_i^*},\\
		\label{repr_factor_full_eq}
		C_R^{\text{eq}} &=& \frac{\frac{c}{d}\frac{k_3}{k_4}}{\left(1+\frac{a}{b}\right)\left(1+\frac{c}{d}\right) + \frac{c}{d}\frac{k_3}{k_4}},
	\end{eqnarray}
	where $u=b+c+d$. $C_R^{\text{neq}}$ and $C_R^{\text{eq}}$ are the expressions for the repression factor~\eqref{repr_factor} for the systems in Figures~\ref{fig:graphs}~(b) and~\ref{fig:graphs}~(c), respectively.
	
	In order to analyze the dependence of $C_R^\text{neq}$ on $k_3$ and $k_4$, we express it as follows:
	\begin{equation} \label{repr_factor_neq}
		C_R^\text{neq} = \frac{A\left(\frac{k_3}{k_4} - \frac{k_1}{k_2}\right)}{B + C\frac{k_3}{k_4} + D\frac{1}{k_4}},
	\end{equation}
	where coefficients $A$, $B$, $C$, and $D$ do not depend of $k_3$ and $k_4$. The critical value $k_4^*$ that  delimits the asymptotic regimes in which $C_R^\text{neq}$ depends on $k_3$ and $k_4$ qualitatively differently can be approximately estimated by equating the last term in the denominator of~\eqref{repr_factor_neq} to the first one:
	\begin{equation}\label{k4_crit}
		k_4^* = \frac{D}{B} = \frac{sd(ak_2 + b(k_1+k_2))}{s(a+b)(c+k_2) + (sb+ac)k_1}.
	\end{equation}
	The asymptotic forms of $C_R^\text{neq}$ are as follows:
	\begin{equation*}
		C_R^\text{neq} \underset{k_4\ll k_4^*}{\sim} \frac{A k_3}{C k_3 + D}, \qquad 
		C_R^\text{neq} \underset{\substack{k_4\gg k_4^*,\\ k_3=O(k_4)}}{\sim} \frac{A\left(\frac{k_3}{k_4} - \frac{k_1}{k_2}\right)}{B + C\frac{k_3}{k_4}}.
	\end{equation*}
	Here, as before, $s = a+b+c+d$.
	
	
	The full expressions for the limits of the repression factor are
	\begin{eqnarray}
		\label{repr_factor_limk3}
		\lim_{k_3\to\infty} C_R^{\text{neq}} &=& \frac{bc(s(c+d+k_2) + uk_1)}{(c+d)(adk_2 + bs(c+k_2) + buk_1)},\\
		\label{repr_factor_limk4}
		\lim_{k_4\to 0} C_R^{\text{neq}} &=&\frac{bck_3(s(c+d+k_2) + uk_1)}{(c+d) \left(dk_2(s(a+b)+ak_3) + bG\right)},
	\end{eqnarray}
	where $G = s(dk_1+ck_3+k_2k_3) + uk_1k_3$.

	\subsection{Experimental data processing}
	The experimental data contain \textit{lacZ} mRNA concentrations for various Giant concentrations, both in arbitrary units. For each genetic construct, we split the \textit{lacZ} mRNA concentrations into 20 bins, with each bin corresponding to a narrow range of Giant concentration, and calculated mean and variance within each bin. However, there is an anomalous jump in the mean and variance in six bins with high Giant concentrations in constructs 4 and 5, which is likely an artifact caused by the small number of data points in those domains. For this reason, we removed from the analysis these bins with the data at $[\text{Gt}] > 0.65$ and $[\text{Gt}] > 0.7$ in constructs 4 and 5, respectively. At the end of this procedure we have 20 bins for each construct 1, 2, and 3 and 14 bins for constructs 4 and 5; the total number of bins $n = 88$.
	
	The Fano factor calculated from the estimated mean and variance was much less than 1 for each bin. This prevents us from using the data in the original form for stochastic modeling, since the simplest deregulated model of mRNA production has the Poisson distribution as the stationary solution, which has a Fano factor equal to 1, while the presence of regulation should lead to the factor values exceeding 1~\cite{Thattai2001}.
	We assumed that the \textit{lacZ} expression in the bin associated with the absence of Giant ($[\text{Gt}]=0$) corresponds to this simplest deregulated case. Therefore, we scaled all \textit{lacZ} expression values from the data with a scaling coefficient that makes the minimal Fano factor (at zero Giant concentration) equal to 1. We used the same scaling coefficient for all constructs. The resulting \textit{lacZ} expression values can be interpreted as estimates of the mRNA copy number, and they have the same order of magnitude as those reported for \textit{lacZ} in the literature~\cite{Xie2008}.

	\subsection{Parameter identification} \label{sec:param_ident}
\begin{figure}
		\centering
		\includegraphics[width=.9\linewidth]{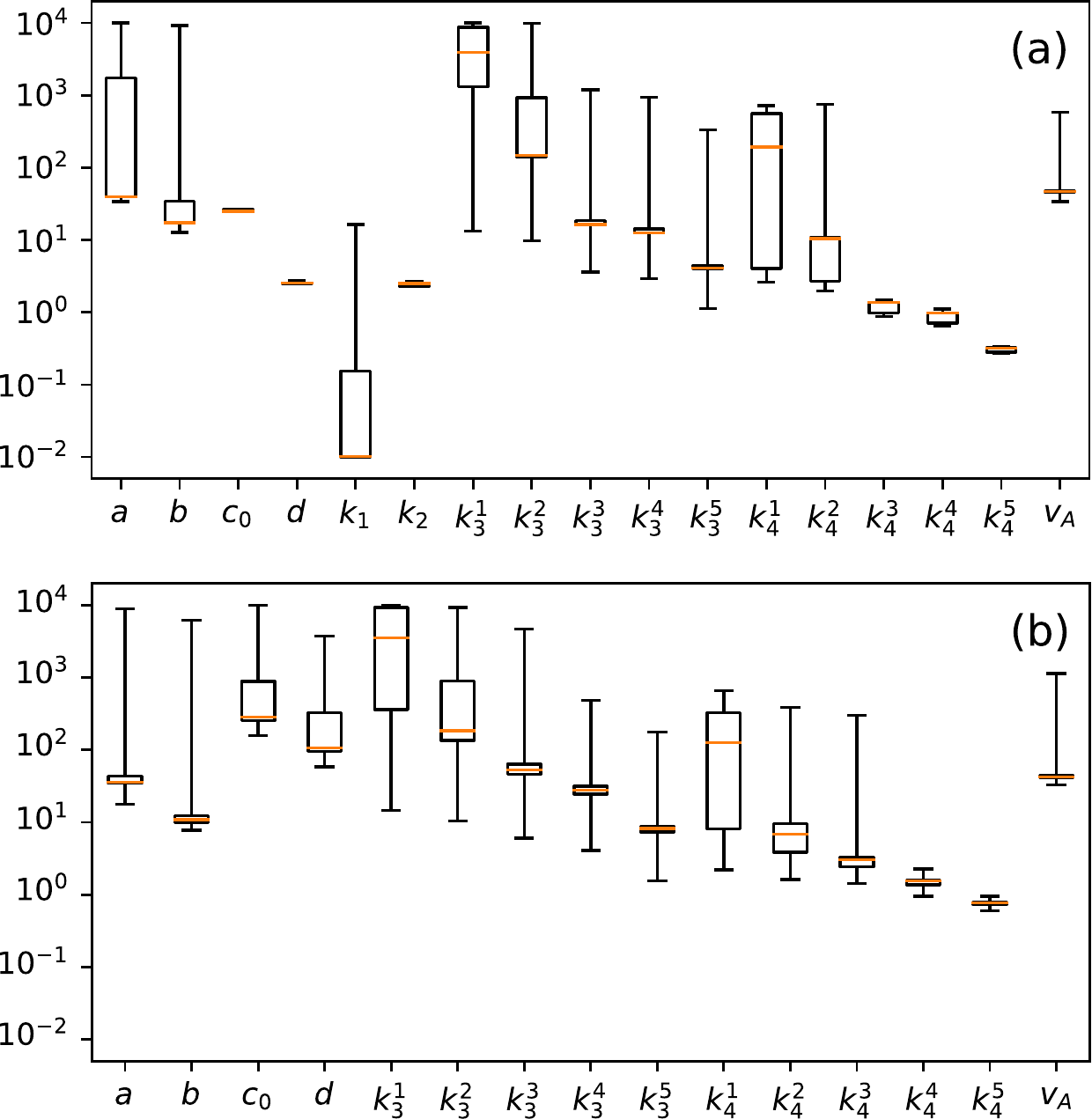}
		\caption{Boxplots for parameter values obtained in 200 optimization runs. Rates $k_3^i$ and $k_4^i$ denote $k_3$ and $k_4$ values for $i$-th construct.}
		\label{fig:6neq_5eq_param_boxplot}
	\end{figure}

	The parameter optimization in the equilibrium and nonequilibrium models produced a wide range of estimated parameter values (Fig.~\ref{fig:6neq_5eq_param_boxplot}).
	As a consequence of parameter nonidentifiability, we observed high correlation between several parameters. In particular, $k_3^i$ and $k_4^i$ exhibit high positive correlation for each construct $i$ {(Fig.~S7~\cite{SupplMat})}. 
	
	To improve the identifiability, we fixed values of $a$ and $b$, which were involved in correlations with other parameters, using estimates for these parameters obtained from the literature. We write $a=[A]/(\tau\gamma)$, where $\tau$ is the typical time required for one activator molecule to find its binding site, and division by $\gamma$ appears because we set $\gamma =1$ in all optimization runs, as discussed in the main text. We use the following estimates: $\tau\approx 100$~s~\cite{Zabet2012}, $[A] \approx 10^4$ is an estimate of protein copy number per blastoderm \textit{Drosophila} nucleus for TFs involved in the segmentation gene network~\cite{Little2011, Zabet2015}, and the \textit{lacZ} mRNA half-life of 3 min. leads to $\gamma \approx 0.4\cdot 10^{-2}\ \text{sec.}^{-1}$~\cite{DONG1995551}. This yields an approximate value $a\approx 10^4\ \text{sec.}^{-1}$, which we use as the fixed value for $a$. We set $b = 10^3\ \text{sec.}^{-1}$ since our optimization runs with free $a$ and $b$ typically produced a one-magnitude difference between these rates. 
	
\begin{figure}
		\centering
		\includegraphics[width=.9\linewidth]{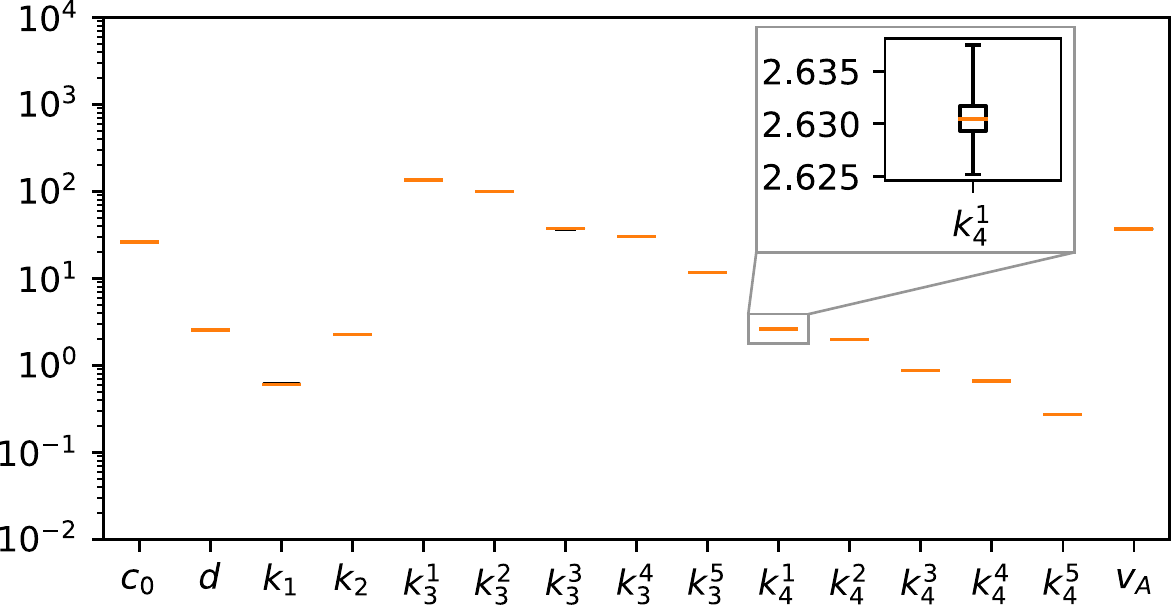}
		\caption{Boxplots for parameter values obtained in 200 optimization runs in the nonequilibrium model with $a = 10^4$ and $b = 10^3$.}
		\label{fig:6neq_fixed_a_b_param_boxplot}
	\end{figure}

	With fixed $a$ and $b$ (in addition to fixed $\gamma$), the optimization resulted in much more precise estimates of the remaining parameters, which essentially solved the non-identifiability problem (Fig.~\ref{fig:6neq_fixed_a_b_param_boxplot}). In particular, $k_3^i$ and $k_4^i$ values also show a small variation for each construct $i$. However, since these rates are allowed to change from construct to construct, we still see the positive correlation between $k_3$ and $k_4$ across the constructs. Therefore, we believe that this correlation is of the same nature as the correlation observed for each construct when these rates were free.
	
	\bibliography{article_en_aps}
	
\end{document}